\documentstyle[manuscript,aps]{revtex}

\begin{document}



\title{Determination of bound state wave functions by a genetic algorithm
  \thanks{supported by DFG and BMFT}}
\author{Christian Winkler\thanks{
        e-mail: {\tt cwinkler@theorie3.physik.uni-erlangen.de}}
        and Hartmut M. Hofmann}
\address{Institut f{\"u}r Theoretische Physik III, Staudtstra\ss{}e 7,
         D 91058 Erlangen, Germany}

\maketitle

\begin{abstract}
We apply a stochastic method of minimizing the ground state energy in
variational calculations of light nuclei using the Refined Resonating
Group Model (RRGM). The method utilizes a bit representation of the
width parameters to be varied. To find the best possible set of width
parameters we use strategies familiar from biological evolution. Very
complicated problems can be solved in this way because the method is
intrinsically parallel.  The algorithm can be used on parallel computers
with any number of processors without any change.
\end{abstract}

\section{INTRODUCTION}

In the framework of the Refined Resonating Group Model \cite{1} bound
states and scattering observables of quite complicated nuclei have been
calculated successfully \cite{2}. In this model a nucleus is decomposed
into clusters.  For details see \cite{1}. Here we give only the
essentials of the wave functions used.  The basic orbital wave function
of a cluster is determined by a Gaussian function. The wave
function of the relative motion again consists of a Gaussian multiplied
by solid spherical harmonics.  The total wave function is constructed
as a linear superposition of antisymmetrized products of orbital and
single particle spin functions. It might be necessary to allow for more
than one width parameter per cluster and to include different
decompositions into clusters, e.g. $^7{\em Li} =\ ^4{\em He} -\ ^3{\em
H} $ and $^6{\em Li} - n$.

However as soon as the nuclei get too complex and many width parameter
are involved it is very difficult to find minima in the space of the
highly nonlinear width parameters. Calculations are very time intensive
since a single evaluation of the binding energy for a fixed set of
width parameters can already take a considerable amount of CPU time.

Traditional methods for finding minima in high dimensional spaces
usually require lots of different function evaluations.  Additionally it
can happen that these methods stop in a local minimum and miss the
global one. A further disadvantage is the fact that all function
evaluations have to be done sequentially because each new test point in
the parameter space depends on the function's value of the previous
points. For this reason the search for the optimal width parameters takes
an enormous time.

One approach to find a smaller binding energy would be simply to extend
the number of linear combinations taken into account
\cite{varga}. However this leads to a very unphysical model where it is
hard to interpret the contribution of each configuration. In addition
to that scattering calculations become almost impossible. Therefore it
would be nice to have a method which is able to find a minimum of the
binding energies in the space of a fixed (but smallest possible) number
of width parameters. Ideally all this should be done as fast as
possible in order to be able to calculate complicated nuclei as well.

\section{OUTLINE OF THE GENETIC ALGORITHM}

Here we briefly summarize the essential points taken from \cite{3}.
Suppose we want to find a minimum in a space spanned by $p$ parameters.
The basic idea of the method is then to have bit representations (called
{\em genes}) of all $p$ optimization parameters which is called an {\em
individual}. Therefore an individual consists of $p$ genes.  Initially
we have to create a certain number $n$ of individuals (a {\em
population} of size $n$) with random genes.  This is called the first
generation.  In an evolution-like method we get from one generation to
the next by selecting two ``good'' individuals.  These will now have the
chance to produce two {\em offsprings} with a crossover-like mechanism
acting on the bit representations.

As we would like to minimize our binding energies with respect to the
width parameters, we chose a binary representation of 16 bit length
(i.e. a 16 bit number) for each parameter. This is sufficiently
precise. The genes are mapped linearly onto the width parameters but
that can easily be changed by using different mappings (e.g. to enhance
the resolution in certain regimes of the parameter space).

The algorithm itself consists of $5$ steps:

\begin{enumerate}

\item
\label{init}
{\em Initialization}.  Create an initial population by randomly
choosing uniformly distributed 16 bit numbers.  Calculate the binding
energies of all members and sort them in ascending order.

\item
\label{choose}
{\em Selection}.  Choose two members of the population according to
their binding energy. Choosing members with lower binding energy should
be more probable. This can be implemented using strategies like roulette
wheel or rank selection\cite{3}.

\item
\label{crossover}
{\em Crossover}.  Use the bit representations of the two selected
members for a crossover to produce two new offsprings \cite{3}.
Crossover can be done in many different ways the most famous
of which is one point crossover (see fig. \ref{crossfig}). However
in most practical applications, as in our case, uniform crossover
(a generalized one point crossover) is used \cite{3}.

\item
\label{mutation}
{\em Mutation}.  Mutate (i.e. invert) each bit in the offsprings with a
given probability. This is done to ensure that the population does not
become degenerate (and hence get stuck in a local minimum) if all bit
representations are similar.

\item
\label{insert}
{\em Insertion}.  Calculate the binding energies of the two offsprings
and insert them into the population. The ``worst'' individuals are
thrown out to have a constant population size.
Return to step \ref{choose}.

\end{enumerate}

The algorithm can terminate after e.g. a fixed number of generations has
been calculated or the mean binding energy of the whole population is
sufficiently close to the lowest binding energy of the population so
that no drastic further change should be expected.

We have to emphasize that the most important process which leads
eventually to convergence is the crossover operation.  The mutation is
only done to ensure diversity in the population.  This can be compared
to the process of biological evolution where it is also thought that
crossover is the most important step.

Usually the most time consuming task is to evaluate the binding energy
for the new parameters in step \ref{insert}. All other tasks are more or
less just bookkeeping. Therefore we note that the whole algorithm is
perfectly well suited for massively parallel computation: each
evaluation of the binding energy can be done on a single processor.

To achieve maximal performance on any parallel computer system our
algorithm proceeds as follows: all available processors are used for
calculating the binding energies of the initial population.  In the
selection process two individuals are selected and their offsprings'
binding energies are evaluated as long as free processors are
available. If no further processor is available the program waits for
the binding energy returned by one of the processors which are sorted
into the current population. Then two new individuals are selected and
their offsprings are sent to the free processors. In this way we try to
assure maximum parallelism.

The time needed for computing a fixed number of generations is therefore
almost {\em inversely proportional} to the number of processors
available. Hence the power of the algorithm grows automatically with
the number of processors available.

\section{RESULTS}

We tried to apply all our considerations to a simple model problem to
see how well it works. For this end we chose the $^7Li$ example. This
is not too simple as $5$ different width parameters are used. On the
other hand calculations do not take too long so that it is easy to
compare the results by the genetic algorithm with those from a
deterministic gradient search.

As a first test we applied the genetic algorithm several times to see if
the method converges on the average.  In all calculations we used a
population size of $50$, a mutation rate of $0.001$ and the number of
total generations was fixed to be $500$.  The results displayed in
figure \ref{genconv} show a reasonable convergence. Of course one has to
do several runs to find reliable results but this is no difference to
the deterministic method where it is possible to become stuck in a local
minimum.

Note that even when the genetic calculation has stabilized basically
still the whole space defined by the mapping of the genes to the width
parameters is used for finding better values for the binding energy. At
this point it is worthwhile stopping the algorithm and to start it
again with a new mapping which takes account of the width parameters
just found.  This can be done several times to enhance the resolution
of the method and to be sure that really the global minimum is found.

The rate of convergence is of course independent of the number of
processors used because all ``administrative'' tasks are done in the
main program which does not depend on the number of processors.

We used the results from the solid line in figure \ref{genconv} again
to estimate how well the genetic algorithm performs compared to a
deterministic search method. Therefore we applied the gradient search
algorithm from the {\tt NAGLIB} \cite{nag} and plotted in figure
\ref{gensmin} the number of necessary steps together with the results
from the genetic algorithm.  Note, however, that the only physical input
to the genetic algorithm was the {\em range} of physically sensible
width parameters whereas the gradient search already needs good {\em
starting values} to yield a reasonable performance. This can be seen by
the much better first energy value in the deterministic method.

However it must be emphasized that the actual time used for getting
the results is the CPU time divided by the number of processors used in
the parallel implementation (except for bookkeeping tasks which can be
totally neglected if the function evaluation takes most of the CPU
time). Therefore it is only of minor interest that the genetic
algorithm needs about twice as many steps (twice the CPU time) as a
deterministic sequential method. This drawback is easily compensated by
the number of available processors.

To get a feeling about the time needed for complicated calculations in
both methods we compare the {\em real time} in figure \ref{gentime}. The
genetic algorithm of course starts later since the initial population
has to be calculated first. After that the genetic algorithm running on
$50$ processors converges dramatically faster than the deterministic
method.

To summarize this section we would like to point out that calculations
of nuclear ground states using parallel genetic algorithms seem to be
very fast and should be prefered against sequential methods. Minimizing
the ground state energy of complicated nuclei might become possible in
much shorter (real) time. The algorithm is very flexible because
parameters like the size of the population, mutation rate, selection
scheme etc. can easily be adjusted to suit the problem.

\section{OUTLOOK}

We have shown that a very easy algorithm ``copied'' from nature can be
used to calculate binding energies and wave functions of rather
complicated nuclei. As this can be done in a highly parallel manner and
is fully scalable many new problems can be solved in this way. The
algorithm is extremely simple and can be generalized to almost any kind
of problem where the determination of an extremum of a function is
involved.

Of course all these computational methods are not restricted to nuclear
physics. Indeed the first test trying to estimate the convergence of
the method was done by calculating the ground state wave function of
the hydrogen atom using the Ritz variational method. So there should
not be any difficulty in using the same methods for quite complicated
quantum chemistry calculations and atomic cluster calculations.

One possible extension of the algorithm is some kind of self adaptive
behaviour, i.e. changing the mapping of the genes to the width
parameters dynamically. This is currently under developed.

\begin{figure}
  \caption{Example of a one point crossover operation between two genes}
  \label{crossfig}
\end{figure}

\begin{figure}
 \caption{Convergence of $3$ different runs of the genetic algorithm
          trying to find the ground state of $^7Li$ (only the binding
          energy of the best individual of the generation is shown).}
 \label{genconv}
\end{figure}

\begin{figure}
 \caption{Convergence of the binding energy as a function of the number
          of steps used in a genetic algorithm (dotted line) and
          a gradient type sequential method (in the gradient method only
          the minimal energy found so far is plotted).}
 \label{gensmin}
\end{figure}

\begin{figure}
 \caption{Comparison of the real time used to find the minimum
          of binding energies using a genetic algorithm running on
          $50$ processors (dotted line)
          and a sequential gradient type search}
 \label{gentime}
\end{figure}

\end{document}